\documentclass[lettersize,journal]{IEEEtran}
\usepackage[nolist]{acronym}
\usepackage{newtxtext, newtxmath}
\usepackage{bm}
\usepackage[pdftex]{graphicx}
\usepackage{gensymb}
\usepackage{cite}
\usepackage{optidef}
\usepackage{bbding}

\usepackage{amsmath,amsfonts}
\usepackage{algorithmic}
\usepackage{algorithm}
\usepackage{array}
\usepackage[caption=false,font=normalsize,labelfont=sf,textfont=sf]{subfig}
\usepackage{textcomp}
\usepackage{stfloats}
\usepackage{url}
\usepackage{verbatim}

\begin{document}

\title{Learning-Based Phase Shift Optimization of Liquid Crystal RIS in Dynamic mmWave Networks}

\author{Le~Hao,~\IEEEmembership{Member,~IEEE,}
        Robin~Neuder,~\IEEEmembership{Member,~IEEE,}
        Mohamadreza~Delbari,~\IEEEmembership{Member,~IEEE,}
        Alejandro~Jiménez-Sáez,~\IEEEmembership{Member,~IEEE,}     
        Vahid~Jamali,~\IEEEmembership{Senior Member,~IEEE,}
        Arash~Asadi,~\IEEEmembership{Member,~IEEE,}
        and~Andrea~Ortiz,~\IEEEmembership{Member,~IEEE}
\thanks{The work of L. Hao and A. Ortiz was supported by the Vienna Science and Technology Fund (WWTF) [Grant ID: 10.47379/VRG23002]. The work of M. Delbari and V. Jamali was supported in part by the LOEWE initiative (Hesse, Germany) within the emergenCITY Centre under Grant LOEWE/1/12/519/03/05.001(0016)/72, and in part by the German Federal Ministry for Research, Technology and Space (BMFTR) under the program of ``Souverän. Digital. Vernetzt.'' joint project Open6GHub plus (Project-ID 16KIS2407). The work of R. Neuder and A. Jiménez-Sáez was supported by the Deutsche Forschungsgemeinschaft (DFG) - Project-ID 287022738 - TRR 196 MARIE within project C09. \textit{(Corresponding author: Andrea~Ortiz)}.}         
\thanks{Le~Hao and Andrea~Ortiz are with the Institute of Telecommunications, Technische Universität Wien (TU Wien), 1040 Vienna, Austria. (e-mail: le.hao@tuwien.ac.at; andrea.ortiz@tuwien.ac.at.}
\thanks{Robin~Neuder and Alejandro~Jiménez-Sáez are with the Institute of Microwave Engineering and Photonics, Technische Universität Darmstadt, 64283 Darmstadt, Germany (e-mail: robin.neuder@tu-darmstadt.de; alejandro.jimenez\_saez@tu-darmstadt.de).}
\thanks{Mohamadreza~Delbari and Vahid~Jamali are with the Resilient Communication Systems Laboratory, Technische Universität Darmstadt, 64283 Darmstadt, Germany (e-mail: mohamadreza.delbari@tu-darmstadt.de; vahid.jamali@tu-darmstadt.de).}
\thanks{Arash~Asadi is with the Embedded Systems Group, Delft University of Technology, 2628 CD Delft, Netherlands (e-mail: a.asadi@tudelft.nl).}}%

\maketitle

\begin{abstract}
To enhance coverage and signal quality in millimeter-wave (mmWave) frequencies, reconfigurable intelligent surfaces (RISs) have emerged as a game-changing solution to manipulate the wireless environment. 
Traditional semiconductor-based RISs face scalability issues due to high power consumption. Meanwhile, liquid crystal-based RISs (LC-RISs) offer energy-efficient and cost-effective operation even for large arrays. However, this promise has a caveat. LC-RISs suffer from long reconfiguration times, on the order of tens of milliseconds, which limits their applicability in dynamic scenarios. To date, prior works have focused on hardware design aspects or static scenarios to address this limitation, but little attention has been paid to optimization solutions for dynamic settings. Our paper fills this gap by proposing a reinforcement learning-based optimization framework to dynamically control the phase shifts of LC-RISs and maximize the data rate of a moving user. Specifically, we propose a Deep Deterministic Policy Gradient (DDPG) algorithm that adapts the LC-RIS phase shifts without requiring perfect channel state information and balances the tradeoff between signal-to-noise ratio (SNR) and configuration time. We validate our approach through high-fidelity ray tracing simulations, leveraging measurement data from an LC-RIS prototype. Our results demonstrate the potential of our solution to bring adaptive control to dynamic LC-RIS-assisted mmWave systems. 

\end{abstract}

\begin{IEEEkeywords}
Liquid crystal, reinforcement learning, RIS, optimization, ray tracing.
\end{IEEEkeywords}

\IEEEpeerreviewmaketitle

\begin{acronym}[DSTTDSGRC]
\setlength{\itemsep}{-3pt}
\acro{AOA}{angle of arrival}
\acro{AOD}{angle of departure}
\acro{AP}{access point}
\acro{BS}{base station}
\acro{CSI}{channel state information}
\acro{DDPG}{deep deterministic policy gradient}
\acro{EM}{electromagnetic}
\acro{LoS}{line-of-sight}
\acro{LC}{liquid crystal} 
\acro{LCD}{liquid crystal display} 
\acro{LTE}{long term evolution}
\acro{MDP}{Markov decision process}
\acro{MEMS}{micro-electro-mechanical systems}
\acro{NLoS}{non-line-of-sight}
\acro{SBR}{shooting and bouncing}
\acro{SISO}{single-input single-output}
\acro{SNR}{signal-to-noise ratio}
\acro{OFDM}{orthogonal frequency-division multiplexing}
\acro{PIN}{positive intrinsic negative}
\acro{RIS}{reconfigurable intelligent surface} 
\acro{RF}{radio frequency} 
\acro{THz}{terahertz} 
\end{acronym}
\section{INTRODUCTION}
\label{sec:Intro}
\IEEEPARstart{M}{illimeter}-wave (mmWave) frequencies (30–300 GHz) offer significantly higher bandwidth and data rates compared to sub-6 GHz bands~\cite{Haider2023, Akyildiz2018}, yet they face severe propagation challenges, such as substantial path loss (20+ dB over sub-6 GHz), blockage sensitivity, and penetration losses higher than 15~dB~\cite{Zhang2022, Kasem2025}. To mitigate these limitations, \acp{RIS} based on technologies such as PIN diodes\cite{Gros2021,Zeng2021}, varactors\cite{Pei2021,Araghi2022,Sievenpiper2003}, RF switches\cite{Rossanese2022}, and MEMS\cite{LiuXuan2022} have emerged as a promising solution allowing passive shaping of the wireless propagation environment to improve coverage and signal strength~\cite{Smart_Radio_Environments, ElMossallamy2020, Jian22}. 
However, these implementations face scalability challenges in terms of power consumption, particularly at mmWave frequencies and when large RIS panels are considered~\cite{Basar2019,Rossanese2022}. For example, prior works show that each PIN diode in a $5\times5$m$^2$ RIS at 10.5 GHz consumes approximately 0.33 mW in the ON state\cite{Tang2021}. This would translate to a power consumption in the order of watts when $10^4$  elements are considered \cite{Jim_nez_S_ez_2022}. In contrast, liquid crystal-based RIS (LC-RIS) technology offers a promising path towards energy-efficient and cost-effective operation for large array configurations. Recent studies estimate that the power consumption of an LC-RIS of the same size with up to $10^6$ elements is around 150mW~\cite{Luis2024}. This dramatic power reduction underscores the potential of LC-RIS for scalable, low-power deployment in next-generation networks.

Despite these substantial advantages, LC-RIS faces a fundamental practical challenge: its configuration time. The physical properties of liquid crystals inherently impose a slow response time, often in the order of tens of milliseconds. This is caused by the fact that the configuration delay scales quadratically with the thickness of the liquid crystal layer~\cite{Neuder2024}. This long reconfiguration time critically limits LC-RIS’s effectiveness in dynamic environments, necessitating novel optimization techniques tailored specifically to this unique constraint.

Existing research on RISs predominantly considers traditional semiconductor-based implementations or static operational scenarios, typically assuming instantaneous reconfiguration~\cite{Jim_nez_S_ez_2022}. This delay-free assumption is however incompatible with real LC-RIS hardware. While prior LC-RIS studies have successfully demonstrated advances in hardware design aimed at minimizing transition delays, algorithmic approaches remain limited, typically relying on simplified optimization techniques or neglecting dynamic communication scenarios altogether~\cite{Delbari2024}. Thus, current solutions do not adequately address the complexity and constraints of realistic LC-RIS-based communication systems, especially under dynamic conditions, see Sec.~\ref{sec:relawork} for a detailed review of related work.
This paper aims to bridge the gaps of LC-RIS optimization from the hardware design to the communication perspective to tackle the main challenges of LC-RISs. The novelty of this paper can be summarized as follows:
\begin{itemize}
    \item We consider for the first time a dynamic LC-RIS-assisted communication system. Specifically, an indoor scenario in which we optimize the phase shifts to maximize the serving time of a moving user. 
    \item Considering that instantaneous perfect \ac{CSI} is difficult to obtain in the mmWave band, we employ a reinforcement learning algorithm for the dynamic optimization of the LC-RIS's phase shifts that does not rely on the availability of perfect CSI.  
    \item We evaluate our proposed approach in a realistic indoor office room with ray tracing simulations, where measurement data of a real LC-RIS prototype is utilized. The results show that we can balance serving time with configuration time without significantly sacrificing the user's \ac{SNR}.  
\end{itemize}

The remainder of this paper is organized as follows. The system model is presented in Section~\ref{sec:systmod}. Section~\ref{sec:probform} introduces the problem formulation. Section~\ref{sec:optprob} explains our proposed solution. In  Section~\ref{sec:simuscene}, we introduce the measurement and emulation with a real LC-RIS. The performance obtained from the algorithm is evaluated in Section~\ref{sec:performance}. In Section~\ref{sec:relawork}, we present some related work to this paper. Finally, Section~\ref{sec:conclusion} concludes the paper.

\section{SYSTEM MODEL}
\label{sec:systmod}
\begin{figure}[t]
    \centering
    \includegraphics[width=0.9\linewidth]{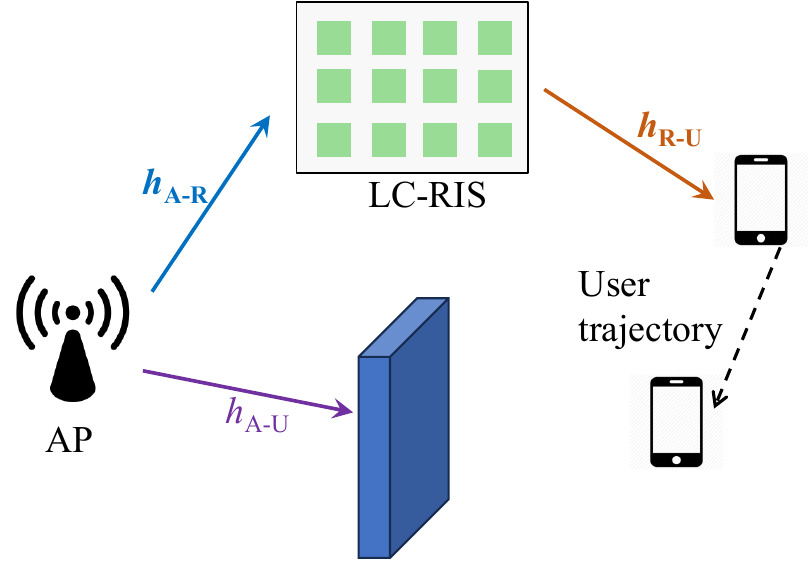}
    \caption{Illustration of an LC-RIS-assisted network scenario.}
    \label{fig:scenario}
\end{figure}

We consider an indoor scenario, such as an office room, conference hall, library, hospital, or home, in which the user positions and channel conditions do not change rapidly, i.e., the user speed is less than 5 m/s. Time is divided into $I$ time slots of equal duration $t_s$. As illustrated in Fig.~\ref{fig:scenario}, the scenario consists of a single \ac{AP}, one mobile user, and one LC-RIS. The AP has one antenna, and the LC-RIS contains $N$ unit cells. Both are located at two different fixed locations. The single-antenna mobile user moves in the room at a constant speed $\nu$ (in m/s). The AP serves the user through the reflection of the LC-RIS. Due to many blockages in the room from walls and furniture, the \ac{LoS} link between the AP and the user is assumed to be blocked. Along with the user movement, the RIS always has \ac{LoS} connections to both the AP and the user. All the $N$ LC-RIS unit cells have a constant amplitude $0\leq\rho\leq1$, but their phases $\omega_n$ can be tuned differently. The reflection coefficient of the $n$-th unit cell is $\gamma_n = \rho e^{j\omega_n}$. For simplicity, in this paper, we assume the LC-RIS does not introduce extra losses and reflects all the signals it receives, i.e., $\rho=1$. The diagonal matrix $\bm{\Omega}\in \mathbb{C}^{N \times N}$ contains $\gamma_n$ as its main diagonal entries. 

We use $ h_{A-U}\in \mathbb{C}^{1 \times 1} $, $\mathbf{h}_{A-R}\in \mathbb{C}^{N \times 1}$, and $\mathbf{h}_{R-U}\in \mathbb{C}^{1 \times N}$ to represent the channels from the AP to the user, from the AP to the LC-RIS, and from the LC-RIS to the user, respectively. In our time-slotted model, the channels $ h_{A-U}$ and $ \mathbf{h}_{R-U}$ vary in every time slot $t_s$, but within each time slot, the channels are constant\footnote{To keep the notation lean, we omit subindices for time slots.}. To ensure the RIS phase shift settings align with the channel state along the user's movement, the LC-RIS phase shifts must be reconfigured every time slot. For convenience, we use $\alpha \in \{A-U, A-R, R-U\}$ to represent the three links, and $K_{\alpha}$ represents the ratio of the received power from the \ac{LoS} and \ac{NLoS} paths for each of the channels. Let the parameter $g_{\alpha}$ denote the large-scale factor of each link, which is obtained from the free space path loss model
\begin{equation}
g_{\alpha} = \frac{\lambda}{4\pi d_{\alpha}},
\end{equation}
where $ d_{\alpha}$ denotes the distance between the network nodes (AP, RIS, user) in the three links, and $\lambda$ is the wavelength.
We characterize the channels with a Rician fading model to account for the presence of a dominant \ac{LoS} component, which can be expressed as
\begin{equation}
\mathbf{h}_{\alpha} = g_{\alpha} \biggl(\sqrt{\frac{K_{\alpha} }{K_{\alpha} +1 }}\mathbf{h}_{\alpha}^{\text{LoS}} + \sqrt{\frac{1}{K_{\alpha} +1 }}\mathbf{h}_{\alpha}^{\text{NLoS}} \biggl).
\end{equation}

When the power of NLoS paths is denoted as $\sigma^2_{\text{NLoS}}$, the \ac{NLoS} component $\mathbf{h}_{\alpha}^{\text{NLoS}} $ is calculated as
\begin{equation}
\mathbf{h}_{\alpha}^{\text{NLoS}} \sim \mathcal{CN}(0, \sigma^2_{\text{NLoS}}).
\end{equation} 
Since the LoS link between the AP and the user is blocked, we set $h_{A-U}^{\text{LoS}}$ as zero so that there are only \ac{NLoS} channels between the AP and the user. Let $(\theta_{\zeta}, \phi_{\zeta})$ stand for the elevation and azimuth angles, respectively, and the steering vectors in terms of the \ac{AOA} or \ac{AOD} from the LC-RIS for the AP-RIS link or RIS-user link are expressed as $\mathbf{a}_{\zeta}$, where $\zeta \in \{A-R, R-U\}$. The \ac{LoS} component of the channel is calculated as
\begin{equation}
\mathbf{h}_{\zeta}^{\text{LoS}} = \mathbf{a}_{\zeta}(\theta_{\zeta}, \phi_{\zeta}). 
\end{equation}
We define the wave number as $\kappa= 2\pi /\lambda $, the space between neighbouring elements in y-direction is $d_y$ and in z-direction is $d_z$, and we can obtain
\begin{equation}
\mathbf{a}_{\zeta}^z = [1, e^{j\kappa d_z \sin(\theta_{\zeta})}, ..., e^{j\kappa d_z (N_z-1) \sin(\theta_{\zeta})}]^\text{T},
\end{equation}
and
\begin{equation}
\mathbf{a}_{\zeta}^y = [1, e^{j\kappa d_y \sin(\phi_{\zeta})\cos(\theta_{\zeta})}, ..., e^{j\kappa d_y (N_y-1) \sin(\phi_{\zeta})\cos(\theta_{\zeta})}]^\text{T}. 
\end{equation}
The steering vector $\mathbf{a}_{\zeta}(\theta_{\zeta}, \phi_{\zeta})$ is defined as
\begin{equation}
\mathbf{a}_{\zeta}(\theta_{\zeta}, \phi_{\zeta}) = \mathbf{a}_{\zeta}^z \otimes \mathbf{a}_{\zeta}^y, 
\end{equation}
with $ \otimes$ representing the Kronecker product. 
In one time slot, the effective channel from the AP to LC-RIS and to the user is calculated as in \cite{Hao2023} as
\begin{equation}
h_\text{eff} = \eta \mathbf{h}_{R-U} \bm{\Omega} \mathbf{h}_{A-R}.
\end{equation}
The term $ \eta = \sqrt{4\pi d_y d_z}/\lambda$ is obtained in  \cite{Hao2023} by comparing the free space path loss model for an RIS-assisted link with ray tracing simulations.  Without $\eta$, the two free space path gains from the AP-RIS and RIS-user links are directly multiplied together, which would ignore the contribution of the RIS.

With a transmitter power $P_{t} $ at the AP, the received power at the user side is calculated as 
\begin{equation}
P_r = P_{t}\left| h_{A-U} + h_\text{eff} \right|^2.
\end{equation}
We denote the noise power as $\sigma^2$ and calculate the user's SNR in one time slot as
\begin{equation}
\text{SNR} = \frac{P_r}{\sigma^2 }.
\end{equation}

Let $t_c$, with $t_c\leq t_s$, be the maximum allowed configuration time of all the LC-RIS elements.
We denote the lower and upper bounds of the phase range for the $n$-th unit cell as $\omega_{\min,n}$ and $\omega_{\max,n}$. The phase $\omega_{\min,n}$ is the achievable phase within one $t_s$ when the previous phase $\omega_{0,n}$ goes in a negative direction (i.e., from $\omega_{0,n}$ to $0$), and $\omega_{\max,n}$ is the maximum phase achievable in a positive direction (i.e., from $\omega_{0,n}$ to $2\pi$ or even higher) in one time slot. We use $\omega_m$ to represent the maximum phase shift that all unit cells can achieve, which is $2\pi$ in our case.  $\omega_{\min,n}$ and $\omega_{\max,n}$ are calculated according to \cite{Delbari2024} as
\begin{equation}
\omega_{\min,n} =\omega_{0,n} e^{\frac{-t_c}{\tau_c^-}},
\end{equation}
and
\begin{equation}
\omega_{\max,n} = \omega_m + (\omega_{0,n} - \omega_m)e^{\frac{-t_c}{\tau_c^+}},
\end{equation}
where $\tau_c^-$ and $\tau_c^+$ are the LC director reorientation time constants. They are set to $\tau_c^- = 29$ ms and $\tau_c^+ = 9$ ms based on the experimental results in \cite{Neuder2024}.
Using the starting phase shift $\omega_{0,n}$ and desired phase shift $\omega_{d,n}$ for the $n$-th unit cell, we  calculate its configuration time $t_{c_n}$ according to \cite{Delbari2025} as 
\begin{equation}
    t_{c_n}=
    \begin{cases}
        \tau_c^+ \ln{\left(\frac{ \omega_m - \omega_{0,n}}{ \omega_m - \omega_{d,n}}\right)} & \text{if } \omega_{d,n} > \omega_{0,n}\\
        \tau_c^- \ln{\left(\frac{ \omega_{0,n}}{ \omega_{d,n}}\right)} & \text{if } \omega_{d,n} < \omega_{0,n}.
    \end{cases}
\end{equation}
The configuration time $t_c$ is then chosen as the maximum over all $t_{c_n}$ as 
\begin{equation}
    t_c = \max_n t_{c_n} \label{equ:tc}
\end{equation}

\section{PROBLEM FORMULATION}
\label{sec:probform}
The main challenge of LC-RIS is the long configuration time. Therefore,  our optimization objective is to maximize the average effective data rate throughout the user's movement by finding a set of RIS phase shifts that minimizes the configuration time in each time slot. We define the serving time of the LC-RIS as $t_k = t_s - t_c$ where $t_c$ is defined in \eqref{equ:tc}, and $t_s$ is the duration of one time slot. After $t_c$, the LC-RIS is in a stable state and serves the user during the remaining time period $t_k$. If the serving time $t_k$ is too short, the RIS spends most of the time tuning its phases and reflects the incident wave in many undesired directions during this transition period~\cite{Alejandro2023}. As a result, the user cannot receive high-quality service when the RIS phase tuning is not completed. 
Let $B$ denote the available bandwidth. We can calculate the effective data rate in one time slot as
\begin{equation}
R = \frac{t_k}{t_s }B\log_2(1+\text{SNR}),
\end{equation}
and formulate our optimization problem as
\begin{maxi!}[3]	  
    	{\bm{\Omega}, t_c} {\mathbb{E}[R] \label{eq:mrt_heu3}}{\label{equ:optProblem}}{}
	\addConstraint{ \left| \gamma_n \right| }{=1, \label{equ:mrt_heu4}}{\quad \forall n=1,...,{N}}
	\addConstraint{0 \leq t_c }{ \leq t_s \label{equ:mrt_heu5}}{}
    \addConstraint{\omega_{\min,n} \leq \omega_n}{\leq \omega_{\max,n},}{\quad \forall n=1,...,{N},}
\end{maxi!}
where the expectation in \eqref{eq:mrt_heu3} is taken with respect to the time horizon $I$.

\section{REINFORCEMENT LEARNING SOLUTION}
\label{sec:optprob}
To solve the optimization problem in \eqref{equ:optProblem}, complete and perfect a-priori knowledge about the user movement
and the channel conditions is required. Such information is not available in real deployments. The channel conditions depend on the user's movements and the RIS is a passive structure which is not able to estimate the \ac{CSI}. Therefore, CSI estimation must be performed at the AP which results in outdated estimated CSI. Solving 
\eqref{equ:optProblem} using outdated \ac{CSI} leads to an inadequate estimation of the system performance.

To overcome this challenge, we consider reinforcement learning to learn how to select the LC-RIS phase shifts. 
The framework of reinforcement learning can be briefly summarized as one agent sequentially interacting with a dynamic environment to learn to make decisions based on its own experiences to maximize a cumulative reward \cite{Puspitasari2023}. 
Note that since the LC-RIS is a passive element, the agent in our case is the AP, i.e., the reinforcement learning algorithm runs in the AP's hardware, and its decisions are the LC-RIS phase shifts. As a result, the AP sequentially selects the phase shifts and informs the LC-RIS about the configuration via existing control channels \cite{Kangwei2025}. 

To apply reinforcement learning we model \eqref{equ:optProblem} as a \ac{MDP}. An MDP is a tuple $(\mathcal{S}, \mathcal{A}, \mathbb{P},
\mathcal{R})$ formed by a set $\mathcal{S}$ of states, a set $\mathcal{A}$ of actions, a state transition probability function $\mathbb{P}$, and a set of rewards $\mathcal{R}.$
In each time slot $i=1,...,I$, the AP observes an state $s_i\in \mathcal{S}$ which is formed by the current phase shifts $\omega$ and the outdated channel state information, i.e., the distances $d_{{A-U}_{i-1}}, d_{{R-U}_{i-1}}$ and the channels $\mathbf{h}_{{A-R}_{i-1}}, \mathbf{h}_{{R-U}_{i-1}}$ in time slot $i-1$. Additionally, to increase the learning speed, we also include in the state the theoretical optimal phase shift $\omega_\text{opt}$ which ignores the configuration time required by the LC-RIS. The value of $\omega_\text{opt}$ is obtained using the outdated CSI as 
\begin{equation}
 \bm{\omega_\text{opt}} = \arg(h_{{A-U}_{i-1}}) - \arg(\mathbf{h}_{{A-R}_{i-1}}^H \odot \mathbf{h}_{{R-U}_{i-1}}).
 \label{equ:optps}
\end{equation}
$s_i$ is then defined as the tuple 
\begin{equation}
\begin{aligned}
s_i = 
& \{ \bm{\omega}, \bm{\omega}_\text{opt}, d_{{A-U}_{i-1}}, d_{{R-U}_{i-1}}, h_{{A-U}_{i-1}},\mathbf{h}_{{A-R}_{i-1}}, \mathbf{h}_{{R-U}_{i-1}} \}.
\label{eq:state}
\end{aligned}
\end{equation}

The action set $\mathcal{A}$ contains the set of possible phase shit configurations $\bm{\omega}$. In each time slot, the AP selects one action $a_i\in \mathcal{A}$, with $a_i=\bm{\omega}$ being the phase shifts to be used in time slot $i$. The state transition probability function $\mathbb{P}$ gives the probability of observing state $s_{i+1}$ after taking action $a_i$ when in state $s_i$. As knowledge about future channel conditions and the user movement is not available, we assume $\mathbb{P}$ is not known at the AP. Finally, $\mathcal{R}$ contains the set of possible rewards $r_i$ the AP will receive when action $a_i$ is taken. In our implementation, we apply reward shaping to improve the flexibility and performance of our proposed solution. Specifically, we do not consider the achieved rate as the reward, as in \eqref{eq:mrt_heu3}. Instead, we define $r_i$ as the linear combination of the achieved SNR and serving time $t_k$ in time slot $i$, i.e., 
\begin{equation}
 r_i = \beta_1\text{SNR} +\beta_2 t_k,
\label{equ:reward}
\end{equation}
where $\beta_1$ and $\beta_2$ are weighting factors to control the trade-off between SNR and serving time.

Since the phase shifts can be configured continuously, the sets $\mathcal{S}$ and $\mathcal{A}$ are infinite \cite{Huang2020}. To handle this continuous state-action space, we consider the \ac{DDPG} algorithm to solve the MDP. 
\ac{DDPG} uses two deep neural networks, one for the actor network that selects the actions $a_i$ and other for the critic network that, based on the selected $a_i$ and the obtained reward $r_i$, evaluates the suitability of the selection. 
As neural networks cannot process complex values, we redefine the state $s_i$ in \eqref{eq:state} as

\begin{equation}
\begin{aligned}
s_i = 
& \{ \bm{\omega}, \bm{\omega}_\text{opt}, d_{{A-U}_{i-1}}, d_{{R-U}_{i-1}}, \Re(h_{{A-U}_{i-1}}),\Im(h_{{A-U}_{i-1}}), \\
 &\Re(\mathbf{h}_{{A-R}_{i-1}}), \Im(\mathbf{h}_{{A-R}_{i-1}}), \Re(\mathbf{h}_{{R-U}_{i-1}}), \Im(\mathbf{h}_{{R-U}_{i-1}}) \},
\end{aligned}
\end{equation}
where $\Re(\cdot)$ and $\Im(\cdot)$ represent the real and imaginary parts of the complex-value channels.

In our implementation, the initial RIS phase shifts are assumed to be random values. Afterwards, the phase shifts of all elements are tuned in each time slot according to the selected action {${a_i=\omega}$} and kept until the next time slot. In the next time slot $i+1$ one, the RIS phase shifts are tuned from the previous $a_i$ values.

\section{MEASUREMENTS AND EMULATION}
\label{sec:simuscene}
\subsection{LC-RIS Measurement}\label{sec:meas}
We consider an LC-RIS containing $30\times25=750$ unit cells. 
Due to hardware limitations \cite{Neuder2024}, the 25 elements in each column have the same phase shifts while tuning, i.e., the phase shifts can only be applied column-wise, resulting in 30 distinct phase shifts. As a consequence, the LC-RIS can only reflect waves toward different azimuth angles, and there is no control over elevated angles\footnote{Despite this hardware limitation, the proposed algorithm can also be applied in two angular dimensions.}. The LC-RIS works at a center frequency of 60 GHz, resulting in a wavelength of $\lambda = 5$ mm. The distance between each element is $0.45\lambda$, and its aperture size\footnote{Note that the minimum required LC-RIS aperture size to deliver the same power as an AP with an unobstructed link can be calculated using \cite[Corollary 2]{Najafi2021}. For our setup, with a minimum distance of $d_{A-U} = 25$~m, a fixed distance of $d_{A-R} = 34.1$~m, and a maximum distance of $d_{R-U} = 15.5$~m, the required aperture size is $D_\text{required}=\frac{\lambda d_{A-R}d_{R-U}}{d_{A-U}}\approx0.106$~m$^2$. Although the area of LC-RIS considered in our setup is smaller than the required aperture size, we compensate for this by using higher transmit power at the AP.} is $D\approx13.5\lambda \times 11\lambda = 37 $~cm$^2$. Therefore, the far-field distance of this LC-RIS starts approximately from $2D/\lambda =1.485 $~m. 

\begin{figure}[t]
    \centering
    \includegraphics[width=0.95\linewidth]{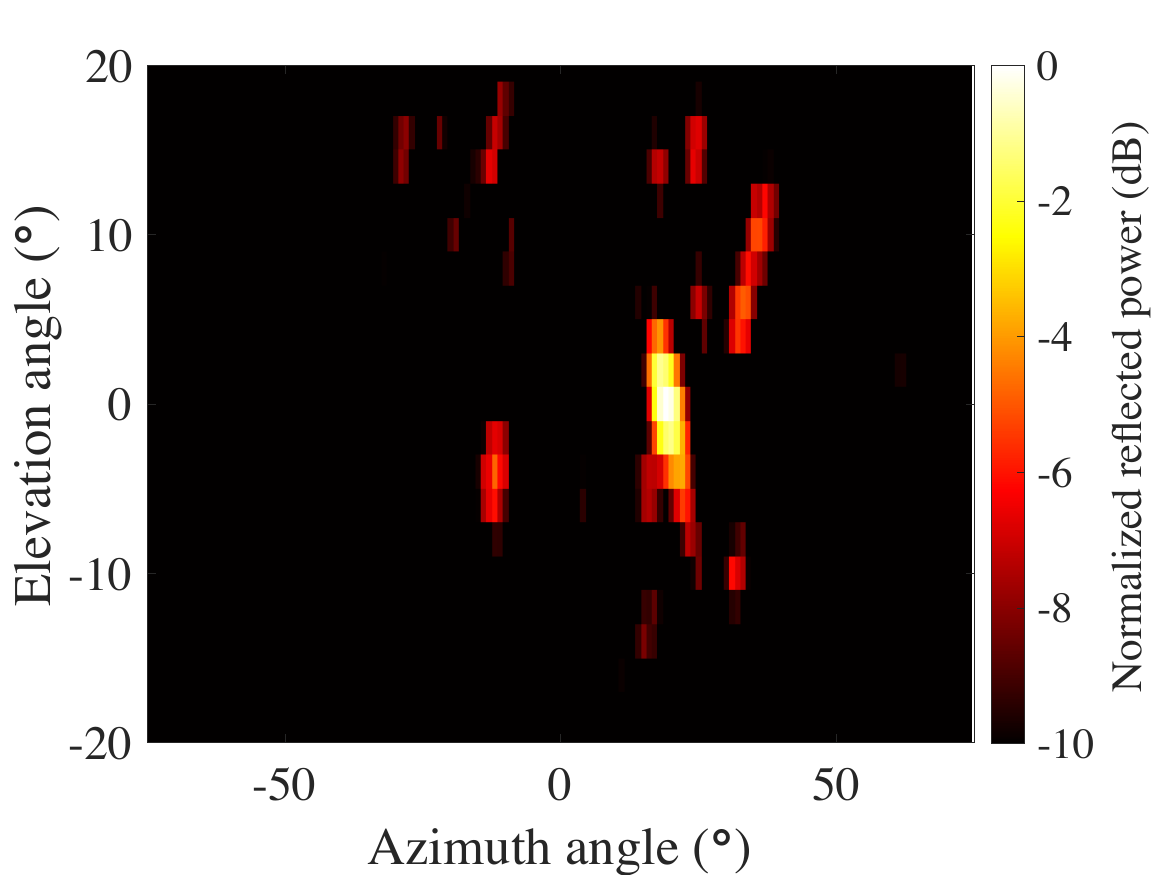}
    \caption{Measured received signal power after sufficient time at 60 GHz and at $20\degree$ steering angle.}
    \label{Fig:meas}
\end{figure}

The LC-RIS response is measured in a lab at 10 different steering angles, from $\phi=-60\degree$ to $\phi=-20\degree$ in steps of $10\degree$, and from $\phi=20\degree$ to $ \phi=60\degree$ in steps of $10\degree$. The incident signal from the Tx antenna impinged on the LC-RIS at $\theta_t=0\degree$(elevation) and $\phi_t=0\degree$ (azimuth) angle. At each RIS steering angle, the reflection responses at azimuth angles of $\phi_r=-74\degree$ to $ \phi_r=74\degree$ with a step of $1\degree$ and at elevation angles of $\theta_r=-20\degree$ to $ \theta_r=20\degree$ with a step of $2\degree$ are measured. Note that $\phi$ is the desired steering direction of the LC-RIS, defined as the azimuth angle from the LC-RIS toward the Rx antenna, and $\phi_r$ is the observation (measurement) azimuth, defined relative to the LC-RIS center, at which the reflection response is recorded. Fig.~\ref{Fig:meas} shows an example of the measured LC-RIS response at the steering angle of $20\degree$ at 60 GHz. 

To apply the phase shift to each LC element, a voltage must be applied. This is achieved using a low-frequency bias voltage, such as a 1~kHz square wave. We consider the DAC60096 EVM with 12 bits from Texas Instruments, which provides a voltage range of $\pm10.5$~V. The measurements were carried out using a PNA-X N5247A from Keysight Technologies \cite{Neuder2024,Delbari2025}. 

\subsection{Ray-Tracing Emulation}
\begin{figure*}[t]
\centering		
\subfloat[\label{Fig:roomside}]{\includegraphics[width=0.9\textwidth]{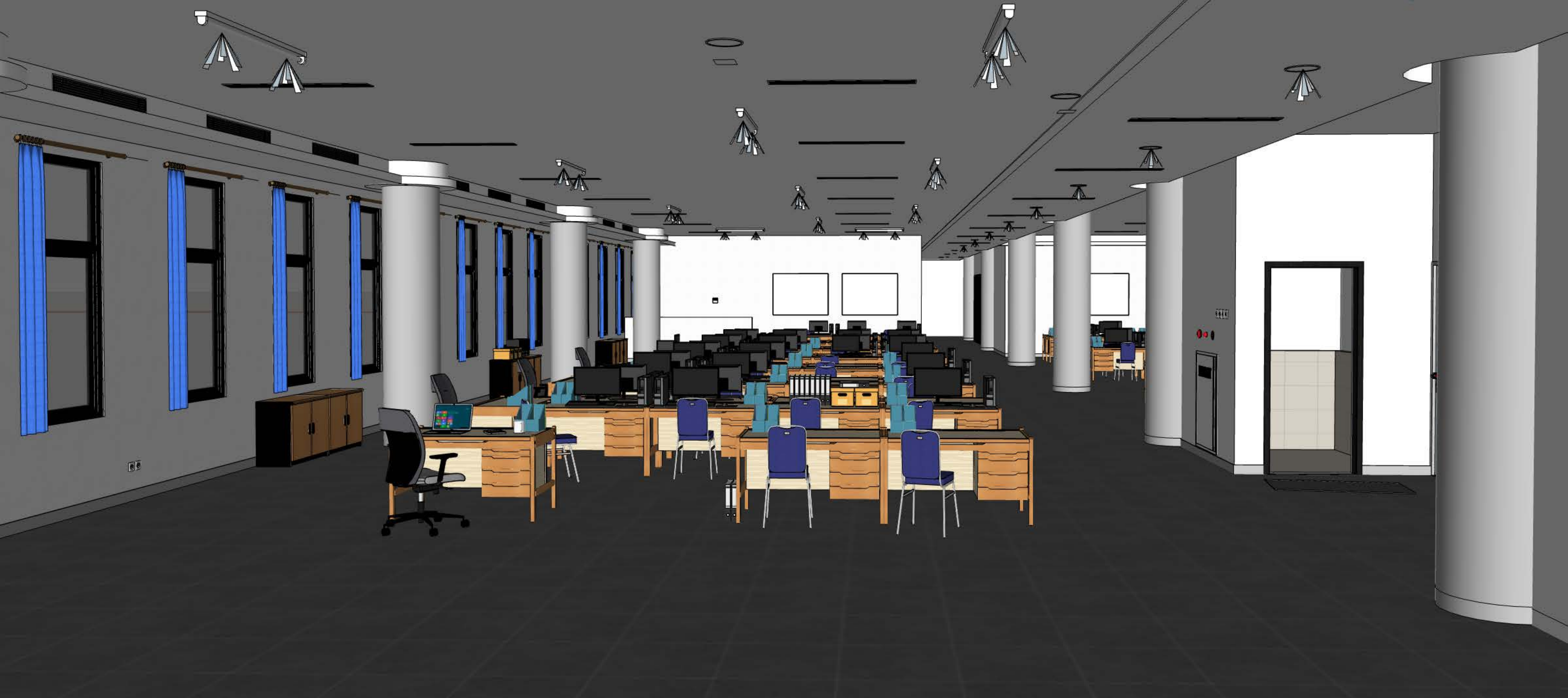}}\\
\subfloat[\label{Fig:roomtop}]{\includegraphics[width=0.9\textwidth]{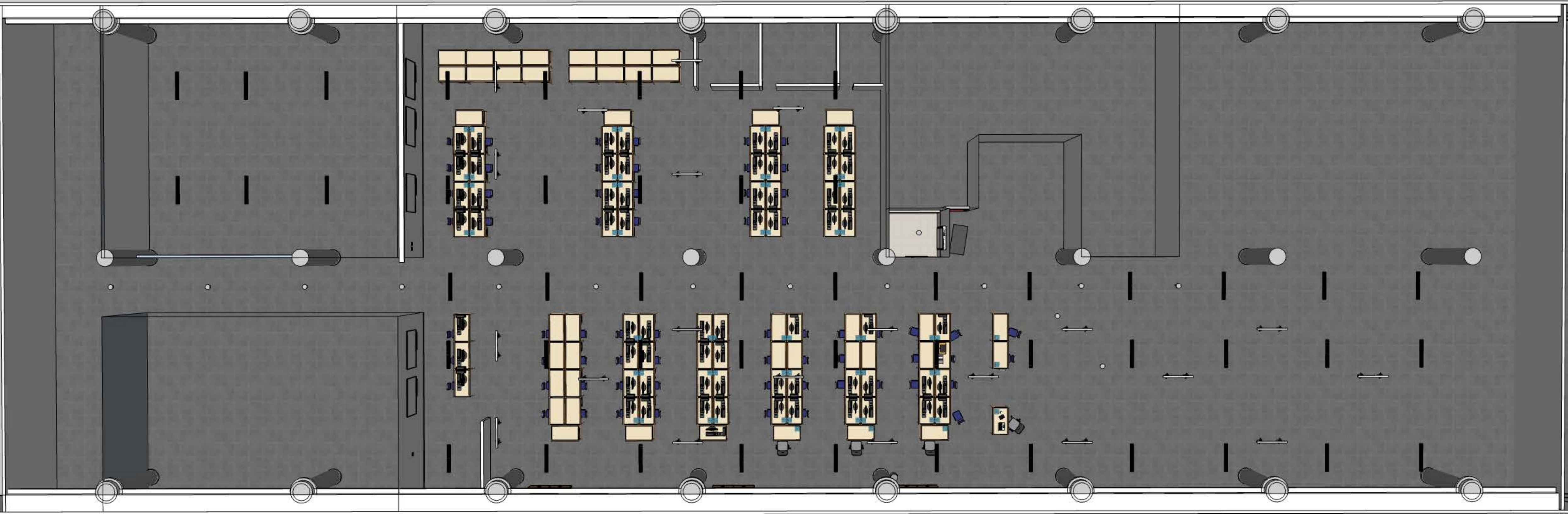}}
\caption{The 3D office room model. (a) The side view of the 3D room. (b) The top view of the 3D room. \label{Fig:RTroom}}
\end{figure*}

\begin{figure*}[t]
    \centering
    \includegraphics[width=0.9\linewidth]{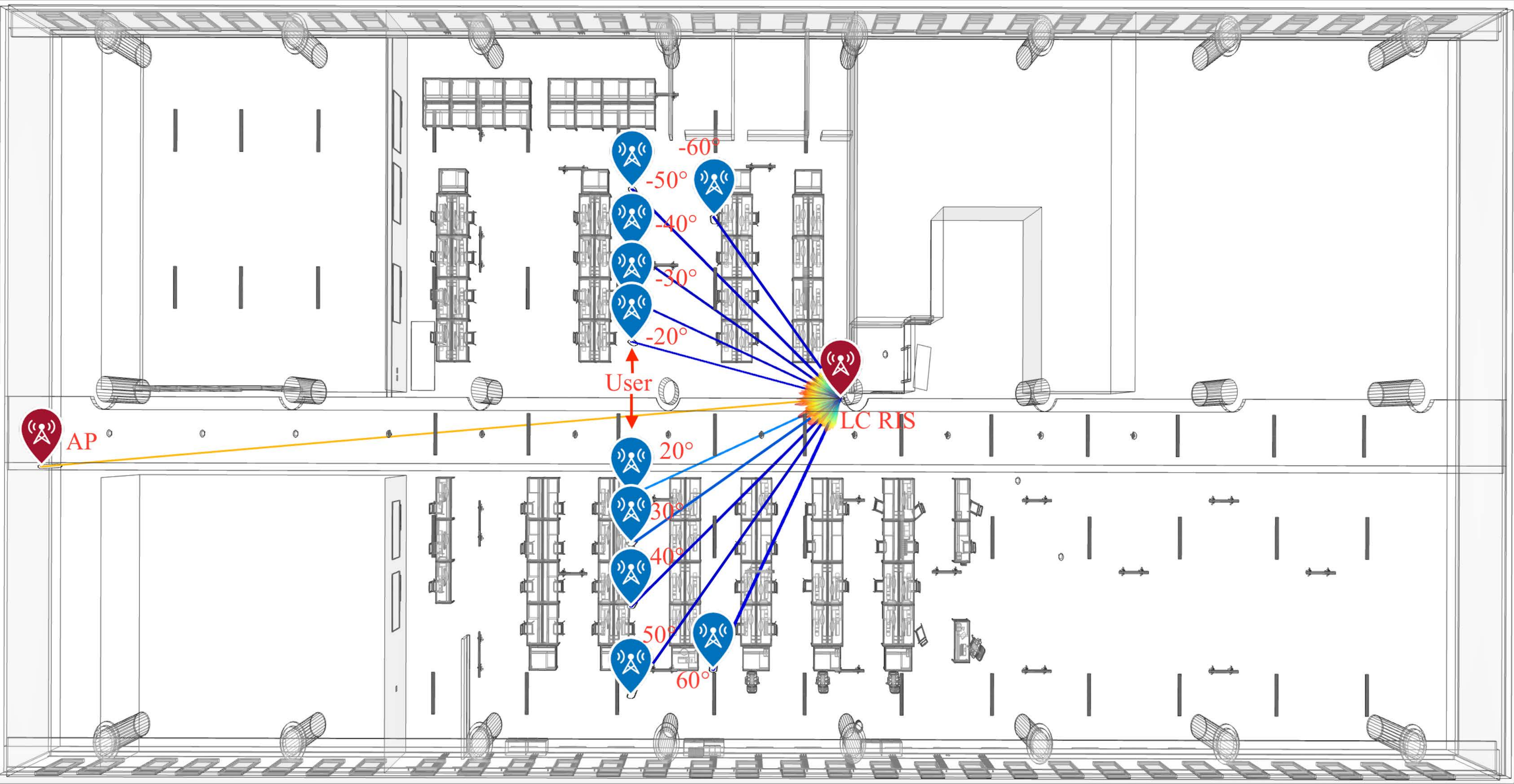}
    \caption{The locations of AP, RIS, and the user in the ray tracing simulation and LoS paths between them.}
    \label{Fig:RT}
\end{figure*}

With the measured LC-RIS scattering pattern, we use a ray tracing tool to predict the performance in more realistic scenarios without performing real measurements. Since this method has been proven to provide accurate and reliable results and is comparable with real measurement results, but eliminates all the measurement costs~\cite{Hao2024}.  
To this aim, we consider an office room as shown in Fig.~\ref{Fig:roomside} and Fig.~\ref{Fig:roomtop}. 
The 3D office room is created in the software SketchUp as an input scenario file for the MATLAB ray tracer. The dimension of length $\times$ width $\times$ height is $63$~m$ \times 36$~m $\times 3$~m. Next, we export the room model to ``.stl'' format and import it into the MATLAB ray tracer\footnote{The MATLAB version up to 2025a does not support multiple material types for an imported ``.stl'' object. Therefore, we set the material type of the whole room object as ``concrete''. The real part of the relative permittivity of concrete is 5.31, and the conductivity is 0.0548 S/m~\cite{mathworks2025}.}. Afterwards, we import the measured LC-RIS response data to the MATLAB ray tracer to model the LC-RIS as an antenna. In addition, we model the Tx and Rx antennas as isotropic ones. The Tx antenna is located at the front of the room. We place the LC-RIS on one of the pillars in the room and place the user at ten different locations, which are at the ten steering angles from the LC-RIS. We assume the user moves from $-60\degree$ to $60\degree$ along the ten location points as shown in Fig.~\ref{Fig:RT}. 

The AP, LC-RIS, and the user are at the same height of 2~m. Fig.~\ref{Fig:RT} shows the locations of the AP, LC-RIS, and user, and the LoS paths between the AP and the LC-RIS, and between the LC-RIS and the user. As described in Section \ref{sec:systmod}, the AP and the LC-RIS have a LoS connection, as do the LC-RIS and the user. However, the AP and the user have no LoS path connections due to the blockage in the meeting room. We run ray tracing simulations with the user at each location and with the scattering pattern of the LC-RIS toward each angle, one by one. To have a better view of the trajectory, we plot all user locations together in Fig.~\ref{Fig:RT}, which contains three straight lines, i.e., from $-60\degree$ to $-50\degree$, from $-50\degree$ to $50\degree$, and from $50\degree$ to $60\degree$. 

For the ray tracer, we use the \ac{SBR} method and set the maximum reflection number to two, three, four, and five. The simulation results show a 0.2 dB received power difference between four and five reflections. As the simulation time is dramatically higher for five reflections, we select four as it provides a better trade-off between accuracy and time. Since the main multipath propagation is from reflection paths in the room, we set the diffraction number to zero. Otherwise, it increases the simulation time significantly but contributes negligible power to the user. The angular separation parameter, i.e., the average number of degrees between launched rays, is set to low. All other simulation parameters are set to the default values. We first model the LC-RIS as a receiving antenna and run a ray tracing simulation from the AP to the LC-RIS with these settings. From this, we obtain the received power at the LC-RIS. Next, we use the received power as transmitted power at the LC-RIS and model the LC-RIS as a transmitting antenna. Finally, we run a ray tracing simulation from the LC-RIS to the user and obtain the user's received power.

\section{PERFORMANCE EVALUATION}
\label{sec:performance}
To validate our solution, we compare our DDPG algorithm with the following approaches using the evaluation parameters given in Table \ref{tab:param}.:\\
\textbf{Optimal}: A theoretical upper bound that assumes instantaneous, unconstrained $2\pi$ optimal phase shifts for all 750 elements. As the configuration time is zero, the optimal approach assumes $t_k=t_s$. This method serves purely as a performance ceiling as such configuration is not physically realizable with an LC-RIS. 
\\
\textbf{Realistic}: A practical benchmark that steers the RIS toward the “Optimal” phase vector but halts tuning after the time slot duration $t_s$ has passed. Thus reflecting true liquid-crystal speed limits. Like in the case of DDPG, the realistic approach considers that only 30 distinct phase shifts can be configured as described in Section \ref{sec:meas}.\\
All the three approaches, i.e., optimal, realistic and DDPG are 
evaluated using the same manufactured LC-RIS scattering measurements described in Section \ref{sec:simuscene}. 
\begin{table}[t]
\fontsize{9pt}{10pt}\selectfont
\caption{PARAMETER SETTINGS}
\label{tab:param}
\centering
\begin{tabular}{c | c }
\hline\hline
 \textbf{Parameter} & \textbf{Values}  \\ \hline
 time slot duration $t_s$ & 10ms\\ \hline
 number of RIS element columns $N_z$ & $30$ \\ \hline
 number of RIS element rows $N_y$ & $25$ \\ \hline
 bandwidth $B$ & $200$~MHz \\ \hline
 user speed $\nu$ & $1.5$ \& 3~m/s \\ \hline
 transmitted power at the AP $P_t$ & $30$~dBW \\ \hline
 noise power $\sigma^2 $ & $-130$~dBW \\ \hline 
 K-factor $K_{A-U}$ & $0$  \\ \hline
 K-factor $K_{A-R}$ & $20$  \\ \hline
 K-factor $K_{R-U}$ & $20$ \\ \hline
 weighting factor $\beta_1$ & $0.2$ \& $0.8$ \\ \hline
 weighting factor $\beta_2$ & $0.8$ \& $0.2$  \\ \hline
 \hline
 learning rate for training actor network & $8.8452e^{-5}$ \\ \hline
 learning rate for training critic network & $1.3876e^{-5}$ \\ \hline
 learning rate for target actor network & $0.0938$ \\ \hline
 learning rate for target critic network & $0.0938$  \\ \hline
 discount factor for reward & $0.9947$ \\ \hline
 buffer size for experience replay & $100000$ \\ \hline
 batch size in the experience & $256$  \\ \hline
 number of time steps for each episode & $19328$ \& $9664$ \\ \hline
 number of episodes & $350$ \\ \hline\hline
\end{tabular}
\end{table}

To evaluate the tradeoff between SNR the serving time, we consider two scenarios:\\
\textbf{Scenario 1:} favors serving time maximization. Therefore,  we set $\beta_1=0.2$ and $\beta_2=0.8$. \\
\textbf{Scenario 2:} favors the maximization of the user's SNR by setting $\beta_1=0.8$ and $\beta_2=0.2$. \\
In each of the scenarios, we consider two user speeds, i.e., 1.5 m/s and 3 m/s. All results from the three approaches are averaged over 350 simulations. 

\subsection{Scenario 1: 
}
\label{sec:scen1}

\begin{figure}[t]
    \centering
    \includegraphics[width=0.9\linewidth]{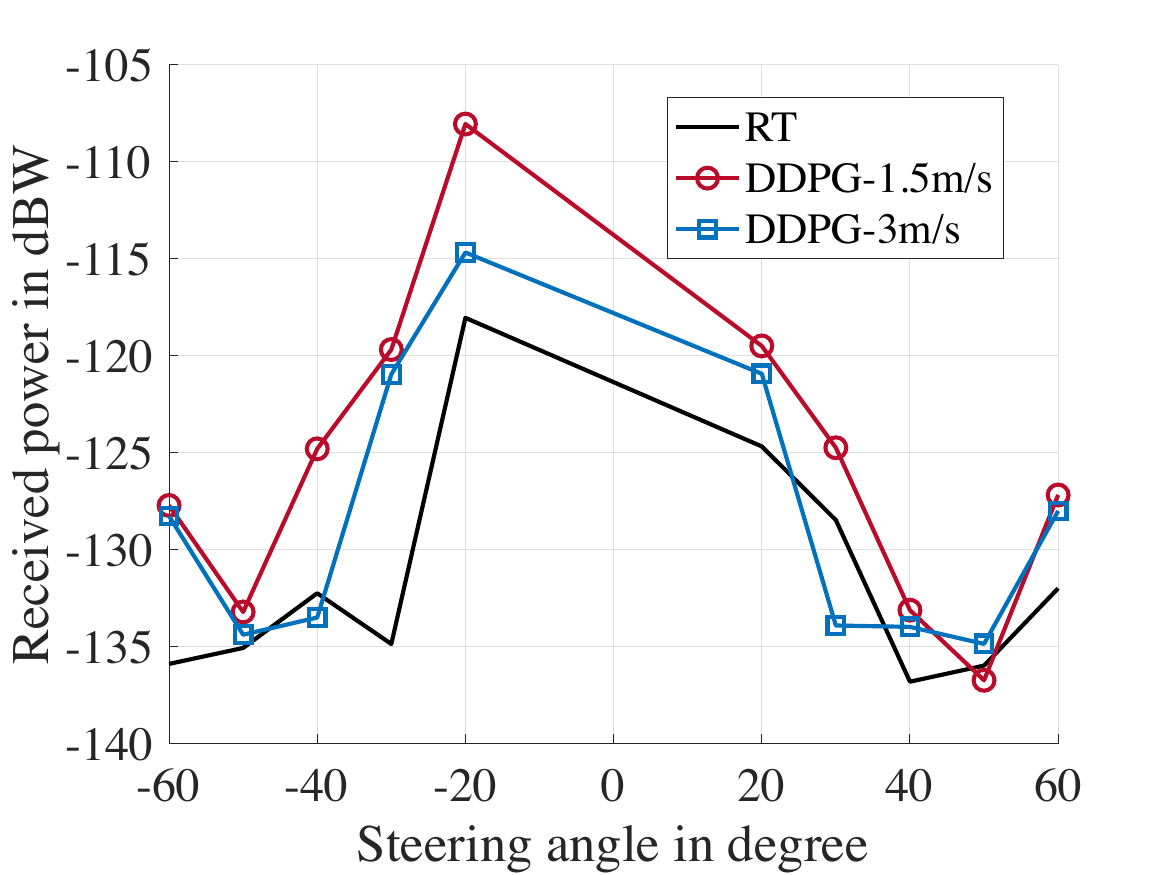}
    \caption{A comparison of the received power between the ray tracing (RT) and the DDPG algorithm simulation with user speeds of 1.5 m/s and 3 m/s.}
    \label{Fig:PWRTRL}
\end{figure}

We first compare the results of the received power obtained from the DDPG algorithm and from the ray tracing emulation to evaluate if the results of the DDPG simulations match those of a real implementation. 
The results are shown in Fig.~\ref{Fig:PWRTRL}. Since the ray tracing simulation only simulates the user at ten fixed locations and there is no movement involved, but the DDPG algorithm simulates a moving user, we pick the results from the DDPG simulation when the user arrives at these ten locations. We observe that the average results from 1.5 m/s are 4 dB higher than at 3 m/s because of the lower impact of the user movement on the outdated CSI. Note that the DDPG algorithm achieves a higher received power than the ray tracing simulations. Specifically, 5.9 dB and 1.9 dB more received power at 1.5 m/s and 3 m/s, respectively. The reason is that the ray tracing uses the manufactured LC-RIS, which has losses in reality, and the ray tracing simulation also includes losses from the reflection materials in the room. However, the DDPG algorithm does not consider any of these losses and therefore leads to a higher received power. In addition, the trends of the received power curves from the DDPG and ray tracing simulations are similar, which indicates that our proposed DDPG algorithm is applicable with a realistic LC-RIS in a realistic scenario.

\begin{figure}[t]
    \centering
    \includegraphics[width=0.95\linewidth]{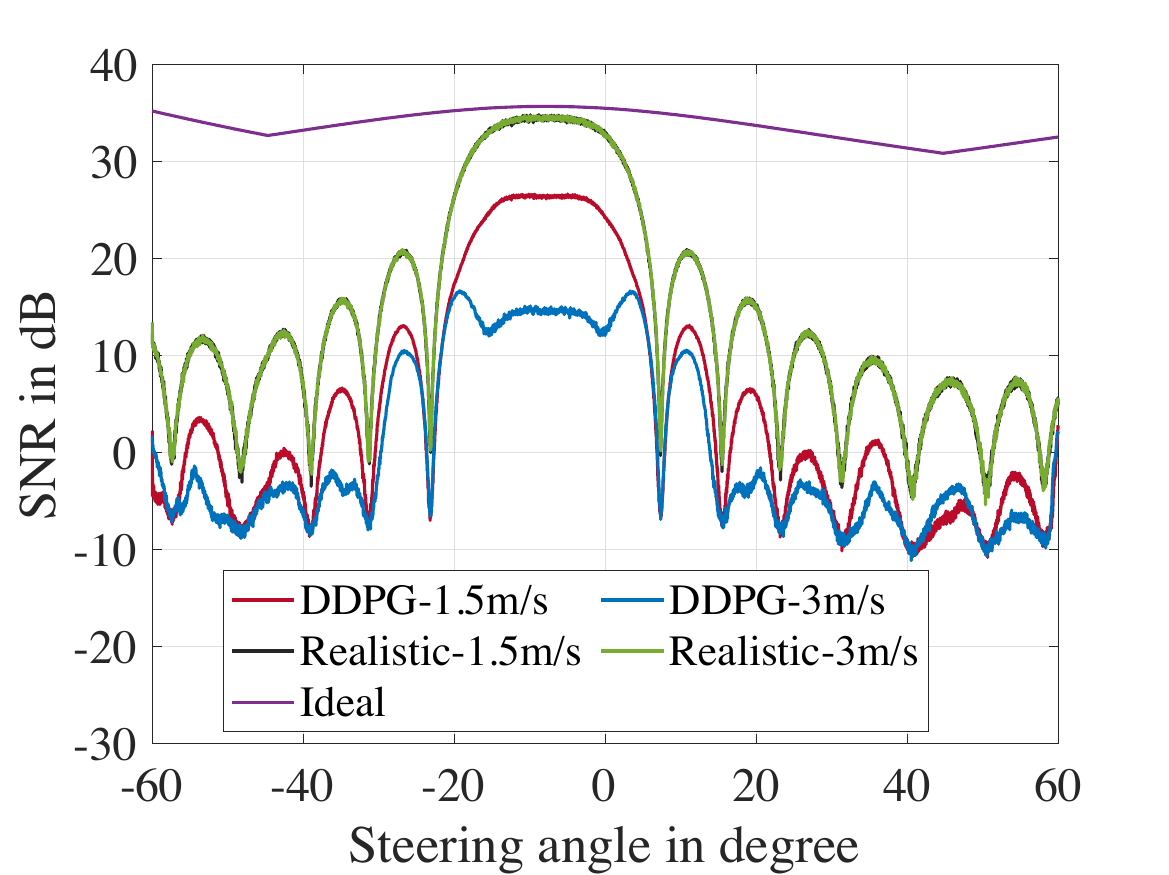}
    \caption{SNR results from DDPG, realistic and ideal cases with user speeds of 1.5 m/s and 3 m/s.}
    \label{Fig:SNR}
\end{figure}

\begin{figure}[t]
    \centering
    \includegraphics[width=0.9\linewidth]{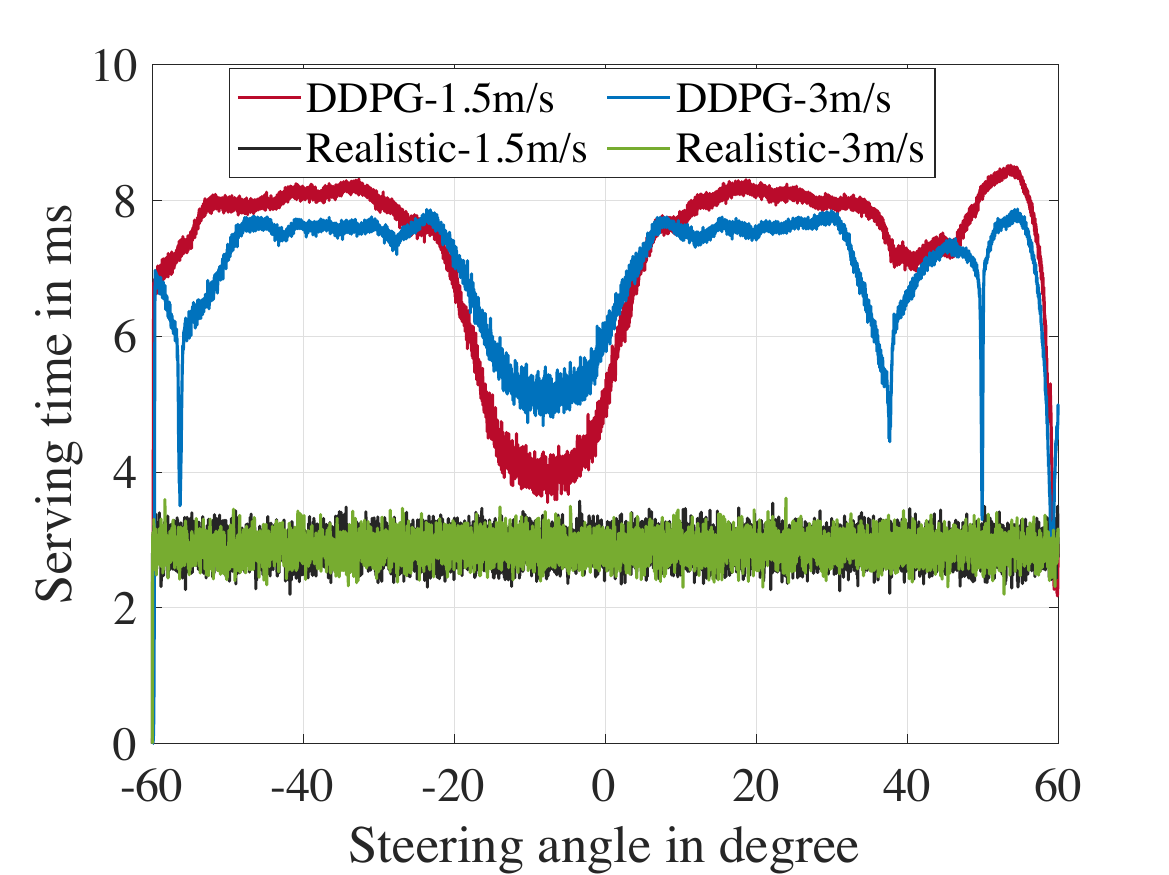}
    \caption{The serving time of the LC-RIS for the user at each location with user speeds of 1.5 m/s and 3 m/s.}
    \label{Fig:servtime}
\end{figure}

\begin{figure}[t]
    \centering
    \includegraphics[width=0.9\linewidth]{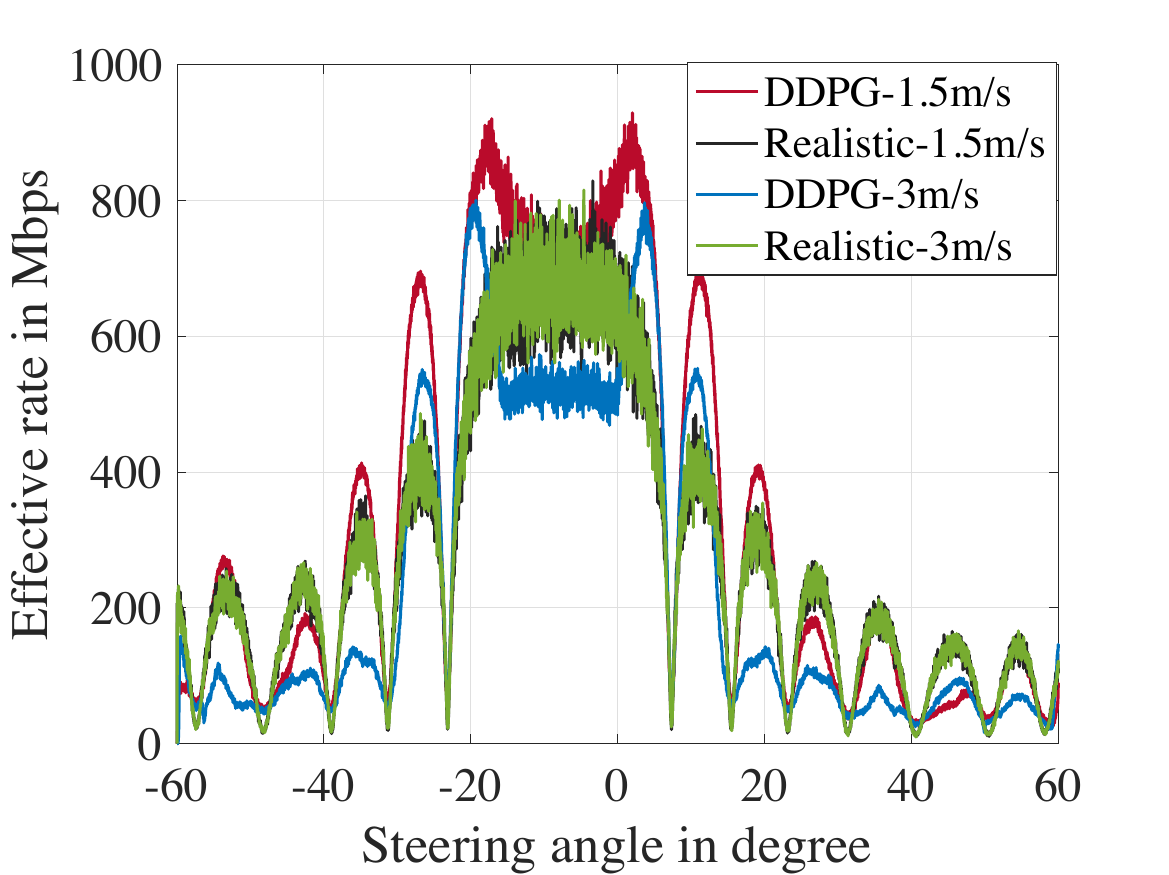}
    \caption{Effective data rate from the DDPG and the realistic case with user speeds of 1.5 m/s and 3 m/s.}
    \label{Fig:rate}
\end{figure}

Next, we compare the SNR results obtained from the DDPG algorithm and from the theoretical calculations, which include the optimal and realistic phase shifts, in Fig.~\ref{Fig:SNR}. The results show that the different speeds only influence the DDPG algorithm, i.e., the SNR for the ideal case with two speeds is the same, and the same for the realistic case. This is because the desired RIS phase shifts are known for each time slot in the realistic and ideal cases, but are unknown for the DDPG algorithm. Thus, the channels for the DDPG case vary in each time slot and with different moving speeds. The SNR from the ideal case is the highest, and the realistic case leads to the second highest SNR since the phase shifts are closer to the optimal ones. The SNR from the DDPG algorithm is the lowest, i.e., it has an average of 9.5 dB (with 1.5 m/s) and 13.5 dB (with 3 m/s) less than the realistic case. This is because the DDPG has the limitation of the configuration time, and only $30$ phases can be changed in a time slot. In addition, the weighting factor $\beta_1$ for the SNR in this case is very low. We can also observe that at a higher speed (3 m/s), outdated CSI causes a larger misalignment between the LC-RIS phase configuration and the actual user location or channel conditions. This misalignment degrades beamforming efficiency, resulting in worse performance (4 dB less) than at the lower speed (1.5 m/s). 

The serving time of the LC-RIS to the user is plotted in Fig.~\ref{Fig:servtime}. The lower speed gives a very similar serving time to a higher speed, i.e., the average serving time from the DDPG algorithm is about 7.1 ms and 6.9 ms for 1.5 m/s and 3 m/s, respectively. This means, the average serving time is reduced only 3\% when the user speed is doubled. The average serving time for the realistic cases of both speeds is about 2.9 ms. The DDPG algorithm gains about 4.2 ms and 4.0 ms of serving time for the user compared to the realistic case, which corresponds to a gain of up to 45\%. 
The effective data rate is plotted in Fig.~\ref{Fig:rate}. With the DDPG algorithm, the effective rate is higher compared to the realistic case, especially in the angle range [$-30\degree ~ 15\degree$]. The average rate from the realistic case is about 277 Mbps, and from the DDPG algorithm is about 328 Mbps and 234Mbps for 1.5 m/s and 3 m/s, respectively. This again proves that a lower speed leads to a higher effective data rate.

\subsection{Scenario 2:}
\label{sec:scen1}

\begin{figure}[t]
    \centering
    \includegraphics[width=0.9\linewidth]{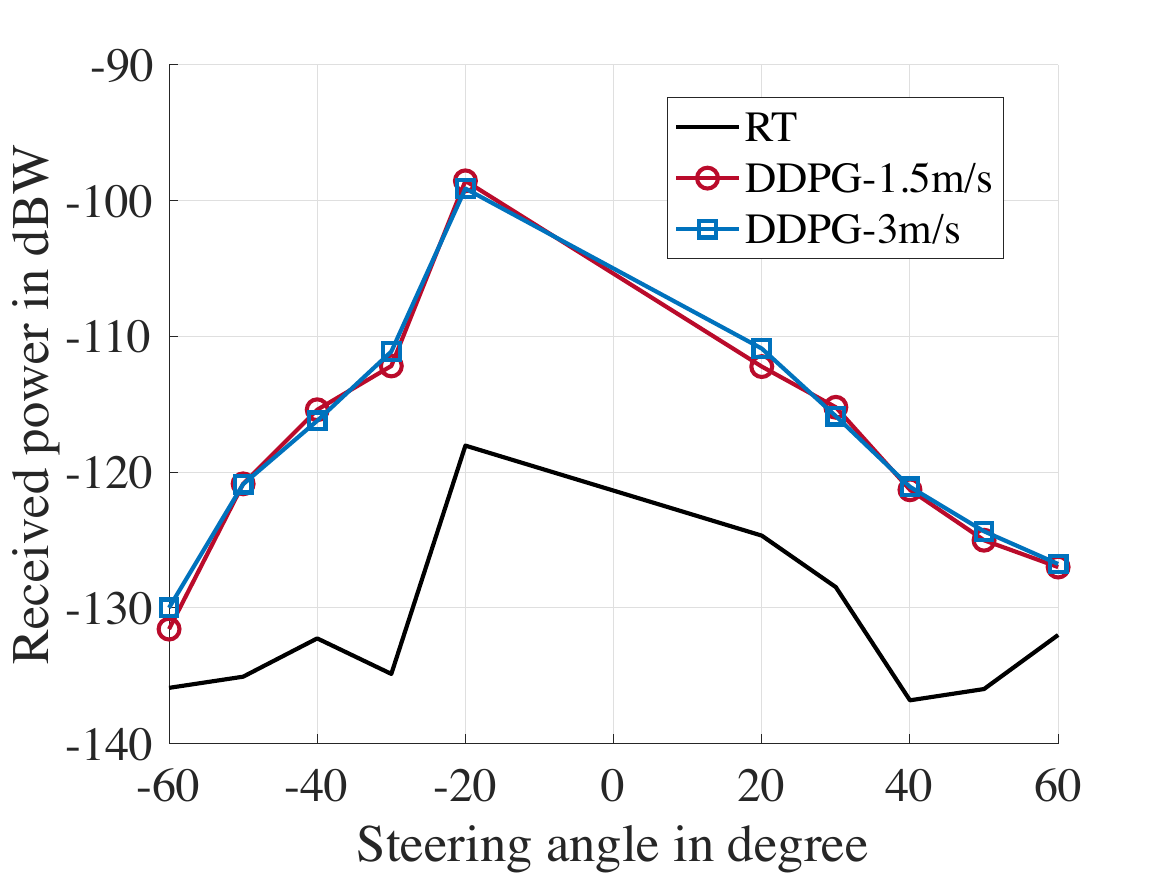}
    \caption{A comparison of the received power between the ray tracing (RT) and the DDPG algorithm simulation with user speeds of 1.5 m/s and 3 m/s. }
    \label{Fig:PWRTRL1}
\end{figure}

\begin{figure}[t]
    \centering
    \includegraphics[width=0.95\linewidth]{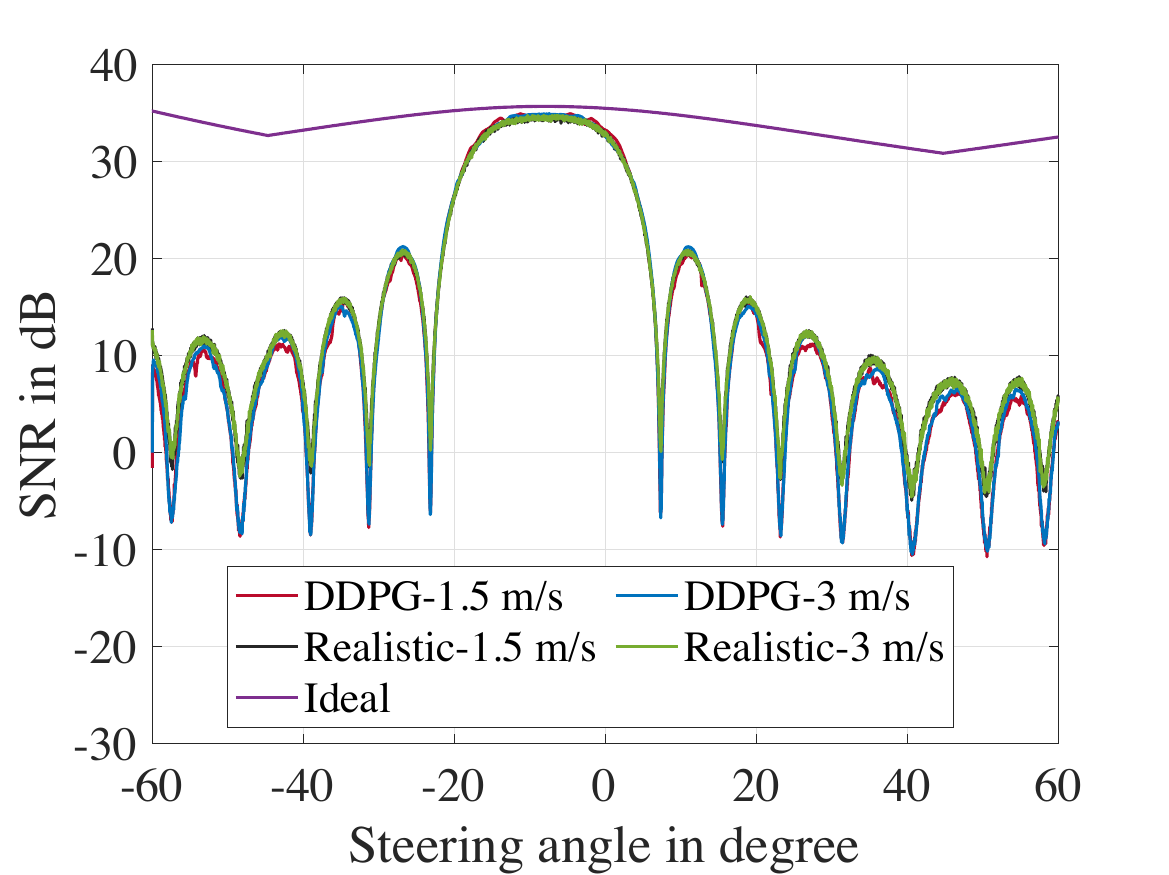}
    \caption{SNR results from DDPG, realistic and ideal cases with user speeds of 1.5 m/s and 3 m/s.}
    \label{Fig:SNR1}
\end{figure}
\begin{figure}[t]
    \centering
    \includegraphics[width=0.9\linewidth]{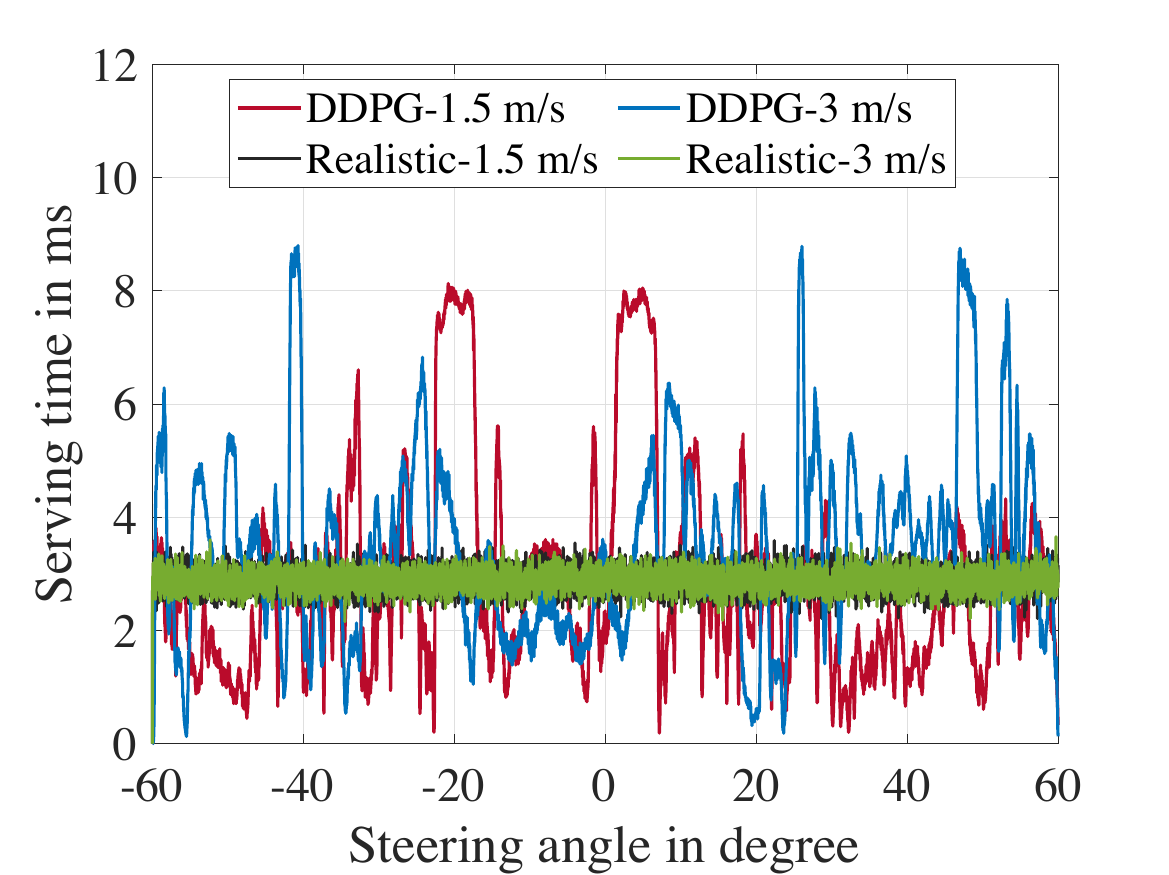}
    \caption{The serving time of the LC-RIS for the user at each location with user speeds of 1.5 m/s and 3 m/s.}
    \label{Fig:servtime1}
\end{figure}
\begin{figure}[t]
    \centering
    \includegraphics[width=0.9\linewidth]{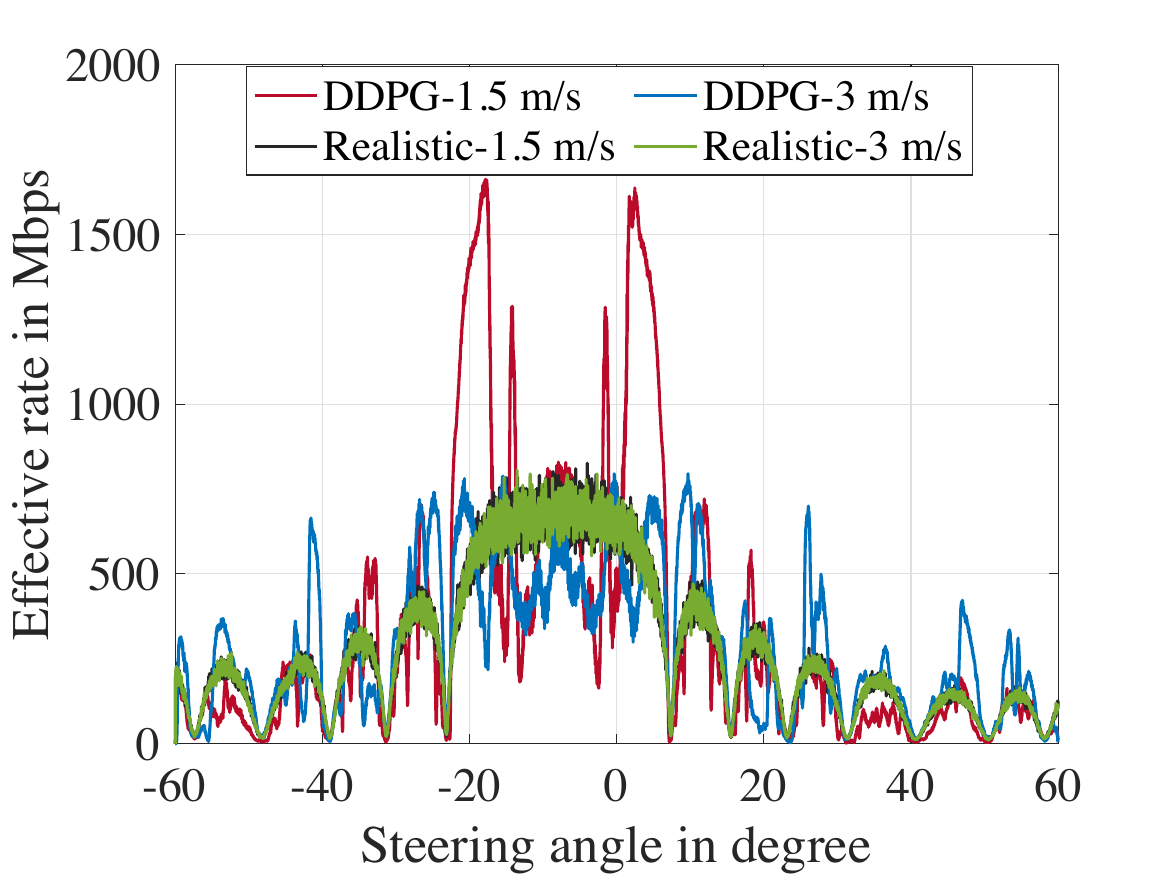}
    \caption{Effective data rate from the DDPG and the realistic case with user speeds of 1.5 m/s and 3 m/s.}
    \label{Fig:rate1}
\end{figure}
Similar to Scenario 1, we first compare the results between ray tracing and DDPG, which are shown in Fig.~\ref{Fig:PWRTRL1}. The DDPG average results from both speeds are almost the same, i.e., around -117.4 dB, and they are about 14 dB higher than the ray tracing results. This is mainly because the SNR is significantly improved in the DDPG algorithm due to the high weighting factor $\beta_1$. 
The SNR results are shown in Fig.~\ref{Fig:SNR1}. The DDPG results are almost the same as the realistic case due to the high weighting factor. We also notice that in the angle range of [$-20\degree ~ 10\degree$], the DDPG, realistic, and ideal cases have the least difference in SNR by 15\% on average. This is because the incident wave from the Tx to the LC-RIS comes at $0\degree$, and this reflection angle range is very close to the incident wave direction. The closer the reflection wave direction is to the incident wave direction, the more power can be received by the user. And the further away reflection angle from the incident angle leads to less received power at the user~\cite{Hao2024}. 

The serving time results are plotted in Fig.~\ref{Fig:servtime1}. The average serving time is about 2.9 ms and 3.4 ms for 1.5 m/s and 3 m/s, respectively. They are about 59\% and 51\% less than in scenario 1. The DDPG algorithm with 1.5 m/s speed gains about 0.6 ms, and the 3 m/s has almost the same serving time as the realistic case. Thus, with this setting of $\beta_1$ and $\beta_2$, we are purely gaining SNR and not gaining serving time. The effective data rate is plotted in Fig.~\ref{Fig:rate1}. The DDPG algorithm gives an average data rate of 327 Mbps for 1.5 m/s and 295 Mbps for 3 m/s, and the realistic case still has 277 Mbps on average. The 1.5 m/s case gains 50 Mbps and the 3 m/s case gains 18 Mbps than the realistic case. In this figure, the average effective data rate with 1.5 m/s is almost the same compared to scenario 1 in Fig.~\ref{Fig:rate}. However, the 3 m/s speed has a slight improvement of 61 Mbps compared to scenario 1. 
These result figures indicate that our DDPG algorithm can achieve a good tradeoff between the serving time and SNR. Depending on the performance preference, one can either sacrifice some SNR to gain more serving time or gain more SNR by reducing some serving time. A lower moving speed is more beneficial for a dynamic wireless network with LC-RISs.

\section{RELATED WORK}
\label{sec:relawork}

LC RIS has been studied in recent years from different contexts, such as hardware design and manufacturing technology, application cases in mmWave and \ac{THz} bands, and optimization algorithm design, each of which has focused on different challenges of LC RISs. In this section, we review the related work in the context of these different perspectives on the main challenges. Furthermore, we shed light on some of the open issues that are not discussed in the state of the art but are tackled in this article.

\textbf{LC RIS design and manufacture.} To the best of our knowledge, most of the existing work on LC RIS focuses on hardware design. The authors in \cite{Luis2024} have explained in detail the architecture, fabrication process, and characterization results of an LC RIS. In addition, an LC RIS architecture with compact defected delay lines has been studied in \cite{Neuder2024}. However, there is still a big room for improvement in terms of component optimization, technological improvements, and low-loss material utilization. Overall, the goal of an optimal LC RIS design is to achieve a good tradeoff between continuous phase tunability, fast configuration time, wide bandwidth, and low loss, but there is a hard limitation on how much the LC RIS hardware can achieve with limited materials and structures. To boost the advantages of an LC RIS, optimizations in the algorithm and communication aspects are needed.

\textbf{Optimization algorithms for LC RISs}
Currently, there are very limited research works on LC RIS optimization problems from the communication point of view. In \cite{Luis2024}, the physical properties of LCs and the beamforming principles of LC RISs have been investigated, and an optimization scheme has been proposed to reduce the transition time. Other optimization algorithms for minimizing the LC RIS configuration time are presented in \cite{Delbari2024} and \cite{Delbari2025}. The proposed algorithms are based on the Lagrange method and the parallel coordinate descent method~\cite{Wright2015}. The existing optimization algorithms include many simplified assumptions to make the problem feasible or easy to solve mathematically, which may not be applicable in real life. The available optimization solutions also do not include the full \ac{CSI}, which would highly degrade the performance in the mmWave band since the channel plays a vital role in the high-frequency range. In addition, the LC RIS analysis in a dynamic scenario is still blank and needs to be investigated. In this context, new optimization algorithms considering these aspects have to be designed. 

\textbf{LC RIS application}
Multiple application cases for LC RISs have been investigated in the literature. For example, \cite{Chandresh2024} presented a 1-bit digital metasurface based on LC for THz wave manipulation. The authors in \cite{Gerardo2013} designed an LC RIS operating in the frequency range of 96-104~GHz and experimentally validated it. An LC RIS operating at 415~GHz has been designed, fabricated, simulated, and tested in \cite{Gerardo2013}, which can achieve wide-angle beam steering. Currently, the LC RIS is mostly used for higher frequencies, such as mmWave and THz communications, and is not suitable for sub-6~GHz communication. Therefore, in this article, we use an LC RIS prototype designed for 60~GHz for performance comparison. Moreover, the existing applications of LC RIS are all in static scenarios, and no study in a dynamic scenario has been published yet. To the best of our knowledge, this is the first work that addresses the performance of an LC RIS in a dynamic scenario. 
\section{CONCLUSION}
\label{sec:conclusion}
This work tackles the main challenge for an LC-RIS, which is the long configuration time. We propose a reinforcement learning algorithm to optimize the phase shifts of an LC-RIS while minimizing the configuration time along the user's movement in an indoor dynamic scenario. We compare the results obtained from the DDPG algorithm and from theoretically calculated optimal and realistic phase shifts. It demonstrates that our proposed algorithm achieves a good tradeoff between the SNR and the configuration time. In addition, we implement a manufactured LC-RIS in a ray tracer to compare the results from ray tracing emulation and the DDPG algorithm, which validates the reliability of our algorithm. The results can be used to estimate the real performance of a real LC-RIS in a realistic room and, therefore, reduce the costs of performing measurements. The optimization for a MIMO network is postponed to future work. Furthermore, it would be interesting to perform measurements with the LC-RIS used in this work in an indoor office room and compare the results between ray tracing, the DDPG algorithm, and measurements in the future.

\ifCLASSOPTIONcaptionsoff
  \newpage
\fi

\bibliographystyle{IEEEtran}
\bibliography{IEEEabrv,Bibliography}

\end{document}